\documentclass{aa}
\usepackage{graphicx}
\usepackage{txfonts}
\input epsf
\usepackage[mathcal]{eucal}
\usepackage{amssymb}
\usepackage{amsmath}
\usepackage{natbib}
\usepackage[T1]{fontenc}

\usepackage{color}

\def\gtrsim{\mathrel{\hbox{\rlap{\hbox{\lower4pt\hbox{$\sim$}}}\hbox{$>$}}}}
\def\ltsim{\mathrel{\hbox{\rlap{\hbox{\lower4pt\hbox{$\sim$}}}\hbox{$<$}}}}

\def\ms {m\,s$^{-1}$}
\def\kms {km\,s$^{-1}$}

\begin{document}

\title{ Pollux: A weak dynamo-driven dipolar magnetic field and implications for its probable planet\thanks{Based on observations  obtained at the the T\'elescope Bernard Lyot (TBL) at Observatoire du Pic du Midi, CNRS/INSU and Universit\'e de Toulouse, France, and the Canada-France-Hawaii Telescope (CFHT), which is operated by the National Research Council of Canada, CNRS/INSU and the University of Hawaii.}}

\author{M. Auri\`ere\inst{1}, P. Petit\inst{2}, P. Mathias\inst{1}, R. Konstantinova-Antova\inst{3}, C. Charbonnel\inst{4,2}, J.-F. Donati\inst{2}, \\ O. Espagnet\inst{1}, C.P. Folsom\inst{2}, T. Roudier\inst{2}, G.A. Wade\inst{5}}
\institute{
IRAP, Universit\'e de Toulouse, CNRS, UPS, CNES, 57 Avenue d'Azereix, 65000 Tarbes, France \\
\email{Philippe.Mathias@irap.omp.eu}
\and
IRAP, Universit\'e de Toulouse, CNRS, UPS, CNES, 14 Avenue Edouard Belin, 31400 Toulouse, France\\
\email{ppetit@irap.omp.eu}
\and
Institute of Astronomy and NAO, Bulgarian Academy of Sciences, 72 Tsarigradsko shose, 1784 Sofia, Bulgaria
\and
Geneva Observatory, University of Geneva, 51 Chemin des Maillettes, 1290 Versoix, Switzerland
\and
Department of Physics, Royal Military College of Canada,
  PO Box 17000, Station 'Forces', Kingston, Ontario, Canada K7K 4B4
}

 \date{Received ??; accepted ??}

\abstract
{Pollux is considered as an archetype of a giant star hosting a planet since its radial velocity (RV) presents very stable sinusoidal variations with a period of about 590\,d. We then discovered a weak magnetic field at its surface using spectropolarimetry, questioning the planetary hypothesis.} {We followed up our investigations on Pollux to characterize its magnetic field and to infer the effects of magnetic activity on the RV variations. }
{We first used ESPaDOnS at CFHT and then Narval at TBL to obtain Stokes $I$ and Stokes $V$ spectra of Pollux to study their variations for a duration of 4.25 years, that is, for more than two periods of the RV variations. We used the least-squares deconvolution (LSD) profiles to measure the longitudinal magnetic field and to perform a Zeeman Doppler imaging (ZDI) investigation.}
{ The longitudinal magnetic field of Pollux is found to vary with a sinusoidal behavior and a period similar to that of the RV variations. 
From the ZDI investigation a rotation period of Pollux is determined to be equal to 660$\pm$15\,days and possibly different than the period of variations of the RV. 
As to the magnetic topology, the poloidal component is dominant and almost purely dipolar with an inclination of 10.5$^\circ$ of the dipole with respect to the rotation axis. The mean strength of the surface magnetic field is 0.44\,G. Pollux is found approximately as active as the Sun observed as a star and this activity could induce moderate RV variations.}
{As to the origin of the magnetic field of Pollux, we favor the hypothesis that it is  maintained through contemporaneous dynamo action. Pollux appears as the representative of a class of slowly rotating and weakly magnetic G-K red giants. To explain the sinusoidal RV variations of Pollux, two scenarios are proposed. If the RV period is different from the rotation period, the observed periodic RV variations are due to the hosted planet and the contribution of Pollux magnetic activity is not significantly detected. In the peculiar case in which the two periods are equal, we cannot discard the possibility that the activity of Pollux could explain the total RV variations and that the planet hypothesis would appear unnecessary. In any case  magnetic activity could contribute significantly to RV variations in some intermediate mass G-K red giants hosting planets, particularly those with small amplitude RV variations. } 
   \authorrunning {M. Auri\`ere et al.}
   \titlerunning {The dynamo-driven weak magnetic field of Pollux}
\maketitle

\section{Introduction}
\object{Pollux} ($\beta$\,Geminorum, HD 62509) is a well-studied K0III giant neighbor of the Sun \citep[e.g.,][]{gra14}. It is considered the archetype of a giant star hosting a planet since it presents periodic sinusoidal radial velocity (RV) variations of about 590\,d period and 40\,\ms\ semi-amplitude, which have been stable for more than 25 years \citep{hce06}. More recently, \citet{awk09} discovered a weak magnetic field at the surface of Pollux, which was the first detection of a class of weakly magnetic G-K giants (Auri\`ere et al. 2015, hereafter AKC; Konstantinova-Antova et al. 2014).  This magnetic field appeared significantly variable, and the magnetic  variations could be correlated with the RV   \citep{awk09}. Following this first result, it is important to determine if the magnetic variations are periodic, possibly resulting from rotation and if the  RV variations could be impacted by, or even reflect, this weak activity. Not only can stellar activity induce jitter, which complicates  planet detection, but the planetary hypothesis has been discarded in some cases of red giant hosts \citep[e.g.,][] {hec18,dls18,rrs19}. We therefore performed a Zeeman long-term monitoring of Pollux  to characterize its magnetic field and to investigate its contribution to the RV measurements. We collected spectropolarimetric data of Pollux using Narval at TBL and ESPaDOnS at CFHT for a duration of 4.25 years, that is, for more than two periods of the RV variations. The results of our investigation were briefly presented by \citet {ake14} and by AKC. In this work, we report the full data, updated investigation, and conclusions, which differ in part from the preliminary findings.

\section{Spectropolarimetric survey of Pollux with Narval and ESPaDOnS}

We observed Pollux on 41 epochs and acquired 265 Stokes V series, first using ESPaDOnS (Donati et al., 2006a) at CFHT in a snapshot program and then its twin Narval at the TBL in a long-term monitoring program, from 2007
September to 2012 February, as described below. All data obtained as part of this project are available through the PolarBase database (Petit et al. 2014). Each instrument consists of a Cassegrain polarimetric module connected by optical fibers to an echelle spectrometer. In polarimetric mode, the instrument simultaneously acquires two orthogonally polarized spectra covering the spectral range from 370 nm to 1000 nm in a single exposure, with a resolving power of about 65000. The observational properties of the instruments, as well as reduction procedures, are the same as described by Auri\`ere et al. (2009) and AKC.

A standard circular polarization observation consists of a series of four sub-exposures between which the 
 half-wave retarders (Fresnel rhombs) are rotated in order to exchange the paths of the orthogonally polarized beams within the whole instrument (and therefore the positions of the two spectra on the CCD), thereby reducing spurious polarization signatures. The extraction of the spectra, including wavelength calibration, correction to the heliocentric frame, and continuum normalization, was performed using Libre-ESpRIT (Donati et al. 1997), which is a dedicated and automatic reduction package installed at CFHT and at TBL. The extracted spectra consist of the normalized Stokes $I$ ($I/I_{\rm c}$) and Stokes $V$ ($V/I_{\rm c}$) parameters as a function of wavelength, along with their associated Stokes $V$ uncertainty $\sigma_V$ (where $I_{\rm c}$ represents the continuum intensity). Also included in the output are the "null polarization" spectra $N$, which are in principle featureless, and therefore serve to diagnose the presence of spurious contributions to the Stokes $V$ spectrum.
 Each spectrum used in this work is of good quality with a peak signal-to-noise ratio (S/N) in Stokes $I$ per 2.6\,\kms\ spectral bin larger than 1000. 

To obtain a high-precision diagnosis of the spectral line circular polarization, least-squares deconvolution 
(LSD; Donati et al. 1997) was applied to each reduced Stokes $I$ and $V$ spectra. We used a solar abundance line mask calculated from ATLAS9 models (Kurucz 1993),  for an effective temperature of 5000\,K, $\log g =3.0$, and a microturbulence of 2.0\,\kms, which is consistent with the physical parameters of Pollux (Hekker and Mel\'endez 2007, Takeda et al. 2008). The mean photospheric profile has an effective Land\'e factor of 1.21 and is centered at 564\,nm. To increase the precision of our measurements, the data were time-averaged each night and, in four cases, for successive nights; each group contained from 4 to 20 Stokes $V$ (and thus also Stokes $I$) series. These data are presented in Table\,1, which provides the log of observations, with the dates, the number of averaged Stokes $V$ (also Stokes $I$) series, the mean S/N for Stokes $V$ profiles, and the mean Heliocentric Julian Date (HJD) of the binned measurement. 

From these mean Stokes profiles we computed the surface-averaged longitudinal magnetic field  $B_\ell$ in gauss, using the first-order moment method (Rees \& Semel 1979), adapted to LSD profiles (Donati et al. 1997, Wade et al. 2000). These measurements of $B_\ell$ are presented in Table\,1 with their 1$\sigma$ error, in gauss, as well as the $B_{\ell,N}$ values computed from the corresponding null polarization spectrum $N$. 

We computed the $S$-index (defined from the Mount Wilson survey; Duncan et al. 1991) for the chromospheric Ca~{\sc ii} H\&K cores to monitor the spectral line activity indicators. These computations are described in detail by AKC. We also computed the heliocentric RV of Pollux from the averaged LSD Stokes $I$ profiles using a Gaussian fit. The RV stability of ESPaDOnS and Narval is about 20-30\,\ms\ (Moutou et al. 2007; AKC).
The averaged value for each date of $S$-index and RV are presented in Table\,1. The data up to 18 March 2009 were already presented by Auri\`ere et al. (2009). All the new observations were made with Narval. The $S$-index measurements were updated to use the procedure described by AKC and those of $B_\ell$ and RV to use the method of Mathias et al. (2018).

\begin{table*}
\caption{Log of observations of Pollux  }          
\label{table:1}   
\centering                         
\begin{tabular}{c c c c c c c c l}     
\hline\hline               
Date          & Number    & S/N         &  HJD      & $B_\ell$ &  $\sigma$ & $B_{\ell,N}$ &$S$-index     & RV \\
              & Stokes $V$\& $I$ & Stokes $V$  &(245 0000+)& (G)     & (G)       &  (G)        & Ca~{\sc ii}  & (km s$^{-1}$) \\  
\hline                      
29 Sep-02 Oct07 &4        & 45794       & 4375.65   &-0.72    & 0.26      &0.43         & 0.123        & 3.646 \\
12 Dec-18 Dec07 &6        & 67831       & 4450.72   &-0.54    & 0.14      &-0.19        & 0.118        & 3.467*   \\
31 Dec 07     &5          & 54334       & 4466.99   &-0.57    & 0.20      &0.05         & 0.120        & 3.536    \\
05 Apr 08     &4          & 76954       & 4562.43   &-0.15    & 0.15      &-0.31        & 0.119        & 3.529    \\
15 Apr 08     &8          & 36901       & 4572.42   &-0.11    & 0.23      &0.09         & 0.120        & 3.506    \\
16 Sep 08     &8          & 49678       & 4726.72   &-0.45    & 0.13      &0.11         & 0.118        & 3.548    \\
20-21 Sep08   &8          & 60062       & 4731.20   &-0.16    & 0.16      &0.32         & 0.119      & 3.572    \\
30 Sep 08     &8          & 76737       & 4740.72   &-0.31    & 0.11      &0.01          & 0.117      & 3.581    \\
21 Dec 08     &8          & 57649       & 4822.62   &-0.32    & 0.15      &-0.18         & 0.118      & 3.615    \\
25 Feb 09     &8          & 74324       & 4888.39   &-0.63    & 0.11      &0.08          & 0.118      & 3.629 \\
12 Mar 09     &8          & 61004       & 4903.43   &-0.71    & 0.14      &-0.03         & 0.116      & 3.643 \\
18 Mar 09     &15         & 62958       & 4909.42   &-0.57   & 0.10       &-0.23         & 0.119      & 3.435* \\
23 Sep 09     &8          & 61466       & 5098.62   &-0.71   & 0.14       &0.06         & 0.118      & 3.536   \\
25-26 Oct09   &12         & 55612       & 5131.21   &-0.49   & 0.12       &-0.23        & 0.117      & 3.538  \\
24 Nov 09     &8          & 58795       & 5160.75   &-0.66   & 0.14       &0.10         & 0.119      & 3.512   \\
15 Dec 09     &8          & 67564       & 5181.76   &-0.31   & 0.12       &-0.11        & 0.118      & 3.476  \\
12 Mar 10     &8          & 54953       & 5268.42   &-0.23   & 0.15       &0.09         & 0.118      & 3.626  \\
10 Apr 10     &8          & 66035       & 5297.33   &-0.24   & 0.13       &-0.03        & 0.118      & 3.613  \\
18 Oct 10     &8          & 65555       & 5488.73   &-0.46   & 0.13       &0.18         & 0.117      & 3.600  \\
18 Nov 10     &13         & 77848       & 5519.70   &-0.51   & 0.08       &-0.02        & 0.116      & 3.620  \\
30 Nov 10     &20         & 27310       & 5531.73   &-0.60   & 0.16       &0.13         & 0.117      & 3.532*  \\
18 Dec 10     &8          & 52935       & 5549.64   &-0.48   & 0.16       &0.16         & 0.117      & 3.645  \\
03 Jan 11     &10         & 62144       & 5565.58   &-0.58   & 0.12       &-0.02        & 0.119      & 3.605  \\
19 Jan 11     &8          & 45256       & 5581.47   &-0.67   & 0.19       &-0.11        & 0.117      & 3.434*  \\
18 Mar 11     &8          & 52109       & 5639.40   &-0.25   & 0.16       &-0.14        & 0.116      & 3.594  \\
25 Sep 11     &8          & 61396       & 5830.70   &-0.54   & 0.14       &0.04         & 0.113      & 3.513  \\
16 Oct 11     &8          & 61620       & 5851.71   &-0.43   & 0.14       &-0.20        & 0.118      & 3.576   \\
23 Nov 11     &8          & 47510       & 5889.76   &-0.71   & 0.18       &-0.22        & 0.120      & 3.510   \\
10 Dec 11     &8          & 62981       & 5906.61   &-0.41   & 0.13       &-0.17        & 0.118      & 3.535   \\
07 Jan 12     &8          & 67918       & 5934.61   &-0.35   & 0.12       &-0.07        & 0.119      & 3.530   \\
08 Feb 12     &8          & 52775       & 5966.54   &-0.33   & 0.16       &-013         & 0.120      & 3.559 \\
\hline                            
\end{tabular}

\tablefoot{Individual columns report date of observation, number of averaged Stokes $V$ \& $I$ series, mean S/N for Stokes $V$ profiles, mean HJD, $B_\ell$ and its error in gauss, $B_{\ell,N}$ corresponding to the null polarization spectrum $N$, the $S$-index for Ca~{\sc ii} H\&K, and the heliocentric RV measured from the LSD Stokes $I$ profile. For details, see Sect.\,2. An asterisk indicates that the measurement is not taken into account in this work.} 
\end{table*} 

\begin{figure}
\centering
\includegraphics[width=9 cm,angle=0] {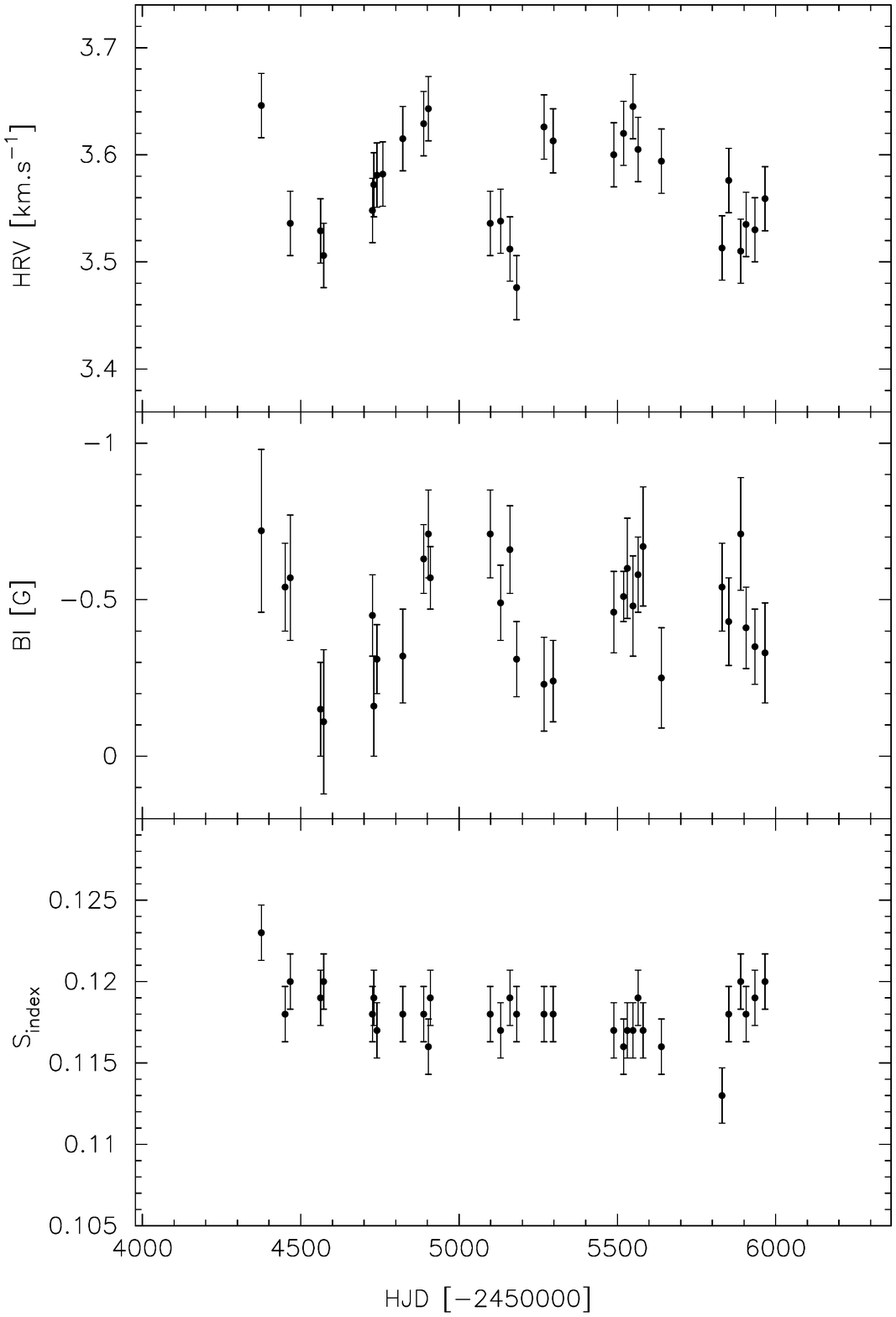} 
\caption{Variations of the heliocentric RV (upper plot), the $B_\ell$ (middle plot), and the  $S$-index (lower plot) of Pollux with HJD.}

\end{figure}

\begin{figure}
\centering
\includegraphics[width=9 cm,angle=0] {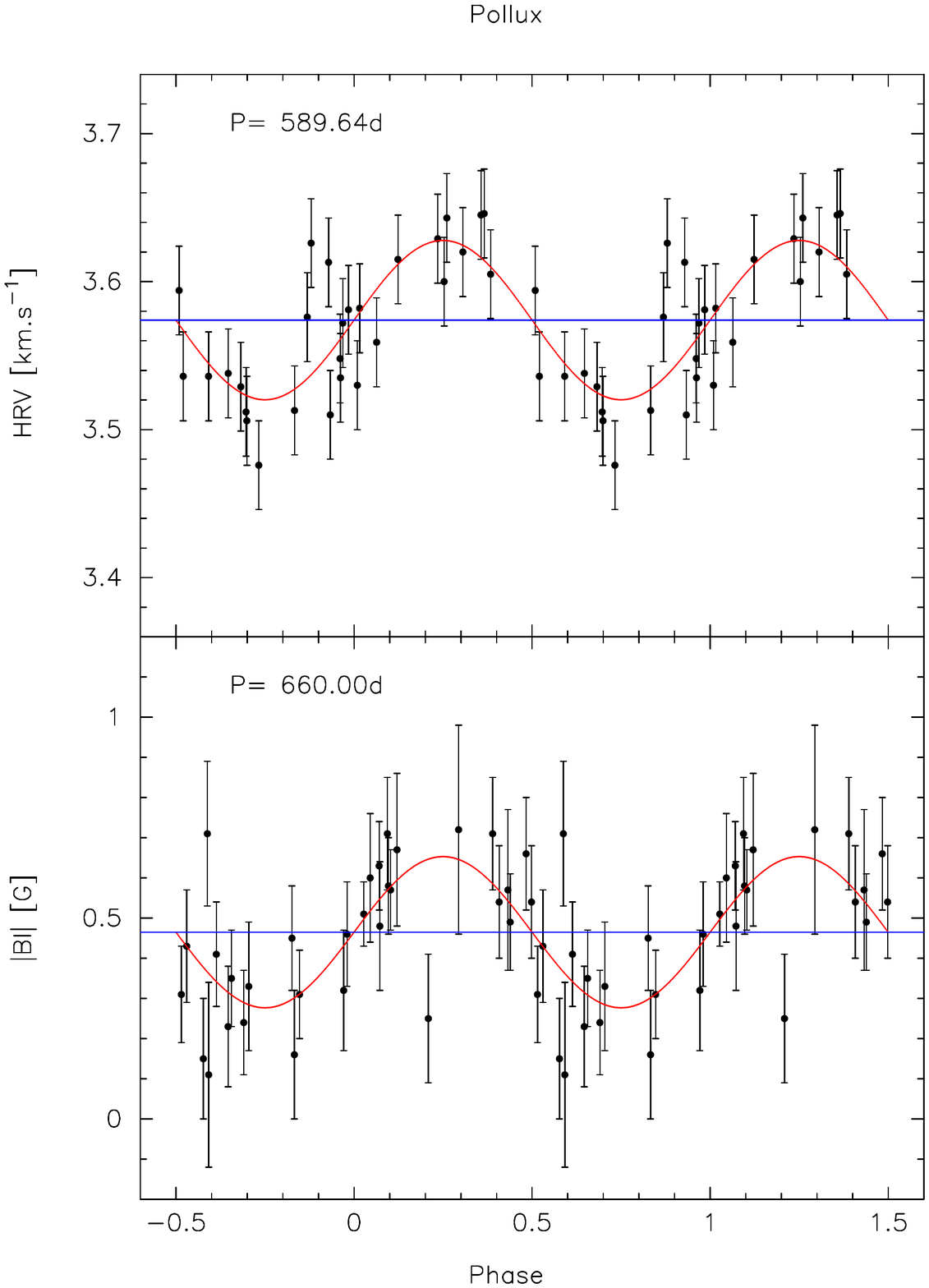} 
\caption{Radial velocity (upper plot) and unsigned $B_\ell$ (lower plot) of Pollux phased with the 589.64\,d and 660\,d periods, respectively. The mean value of RV and  $|B_\ell|$ are shown. Two periods are plotted for clarity.}

\end{figure}  

\section {Variations of the RV, of the longitudinal magnetic field $B_l$, and the $S$-index of Pollux}

Extensive studies of the RV of Pollux have been performed following the discovery of a RV variation with a total amplitude of about 100\,\ms\ (Walker et al. 1989) and a period of about 580\,d (Larson et al. 1993, Hatzes \& Cochran 1993). To interpret these variations, the planetary companion hypothesis is now generally accepted (Hatzes et al. 2006, Reffert et al. 2006, Han et al. 2008). These three previous investigations give periods for the RV variations near 590\,d. Table\,2 shows the period, its uncertainty, and the span time of the observations for each reference. Baklanova et al. (2011) used all the RV from these studies to derive a new period of 592.9$\pm$0.6\,d and also provided a few $B_\ell$ measurements. We used 589.64\,d as the RV period of Pollux (Hatzes et al. 2006) since this period comes from the longest span of observations with homogeneously reduced data leading to the most accurate period determination. 

Figure\,1  (upper plot) shows the variations of RV with HJD. The error bars are fixed to 30\,\ms\ (see Sect.\,2). For four dates, the RV values were outliers from the sine trend (being generally significantly smaller) and were discarded. These values are denoted by an asterisk in Table\,1 and they are not presented in the plot. Figure\,1 (middle plot) shows the variations of $B_\ell$ with HJD (only our homogeneous $B_\ell$ measurements were used). The error bars are those presented in Table\,1. The lower plot of Fig.\,1 presents the variations of the $S$-index. Error bars of 0.002, corresponding to the largest deviation from the mean (apart for two cases), are shown.

Figure\,1 clearly confirms the trend shown by Auri\`ere et al. (2009), in which $B_\ell$ appears to mimic the variations of RV. Compared to the observed variations, the error bars on $B_\ell$ are however larger than for RV, leading to a noisier plot. On the other hand, the $S$-index (lower plot) does not show any clear trend, remaining at a very low level; this behavior is typical of a very weakly active star as described by AKC. 

Our Zeeman Doppler imaging (ZDI) study of Pollux using the Stokes $V$ data,  presented in Sect.\,4, determines a rotation period $P_{\rm rot}$ of 660$\pm$15\,d (statistical error), which is different from the 589.64\,d period of the RV variations at the 4$\sigma$ level. However we do not exclude that the two periods could be equal. This possibility is discussed in Sect.\,4 and Sect.\,6.

We carried out a classical Fourier analysis of the $B_\ell$ and RV data and found  possible periods of 635\,d and 661\,d, but the errors are about 40\,d and 27\,d,  respectively, because of our limited sample and poor relative precision on $B_\ell$ and RV. For example, our relative error is about 30\,\ms\ on RV measurements, when it is a few \ms\ for the works referenced in Table\,2. These periods are therefore compatible with the two other periods (589.64\,d and 660\,d). We consider in this work  that the $P_{\rm rot}$ of 660\,d is responsible for the $B_\ell$ variations. The similarity of the curves of variations in RV and $B_\ell$ is due to either chance if the two periods are really different, or to a true physical relationship if the two periods are equal. 

Figure\,2 then shows the RV (upper plot) and the $B_\ell$ (lower plot) measurements of Pollux phased with the 589.64\,d and 660\,d periods, respectively.  Two periods are shown for clarity. Each sinusoid is fitted with the least-squares method on each data set. The periods are fixed, while the semi-amplitude, the mean value, and the phase origin (corresponding to the mean value of RV and $B_\ell$) are  the only free parameters. This gives  amplitudes of variations  of 53.8\,\ms\ and 0.19\,G, respectively.  For these sine fits, the frequencies corresponding to  the two periods account for  58$\%$ and 41$\%$ of the fraction of the variance for  RV and $B_\ell$, respectively.  The difference between phase origins in RV and $B_\ell$ is about 92 days; if this is due to the difference in periods this value should vary with time, increasing by about this difference after each rotation of Pollux or revolution of the planet. Because of the large errors on $B_\ell$ and RV, as described above, we were unable to confirm a possible variation of this difference between phase origins during the span of our observations.

\begin{table}
\caption{Observed periods on  Pollux  and parameters for a hosted planet}          
\label{table:1}   
\centering                         
\begin{tabular}{l c c c c c c}     
\hline\hline 
Reference    & Period & Error  & Span & $ m_2 \sin i$ & $a$ \\
             & (Day)  & (Day)  &(Year)& ($ M_{\rm Jup}$) & (AU) \\
\hline
Hatzes 2006  & 589.64 & 0.81   & 25.1 & 2.3        & 1.64 \\
Reffert 2006 & 589.7  & 3.5    & 3.5  & 2.7        & 1.69  \\
Han 2008     & 596.6  & 2.26   & 4.5  & 2.47        & 1.66 \\
\hline
ZDI          & 660    & 15     & 4.25 \\ 
$B_\ell$      & 635    & 40     & 4.25 \\
RV        & 661    & 27     & 4.25 \\
\hline                           
\end{tabular}

\tablefoot{Numbers from the 3 first references are used in Sect.\,3 and Sect.\,6. The quantity  $ m_2 \sin i$ is given for a mass of Pollux of 1.7 $M_{\odot}$ as used in these 3 references. The ZDI work is described in Sect.\,4. The $B_\ell$ and RV work is presented in Sect.\,3.}
\end{table}

\section {Zeeman Doppler imaging of Pollux}

To fully exploit the Zeeman information included in our Stokes V data, we employed the ZDI method, using the new implementation of Folsom et al. (2018). This Python version implements the same physical model and inversion method as the code detailed in Donati et al. (2006b). The two ZDI codes were extensively tested to ensure they produced identical results for identical input parameters. The ZDI method has already been successfully used to get magnetic field distribution at the surface of five red giant stars, including both fast rotators (\object{V390\,Aur}; Konstantinova-Antova et al. 2012; \object{37\,Com}, Tsvetkova et al. 2017) and Ap-star descendant candidates (\object{EK\,Eri}; Auri\`ere et al. 2011; \object{$\beta$\,Cet}; Tsvetkova et al. 2013; \object{OU\,And}; Borisova et al. 2016). The peculiarity of Pollux with respect to these five red giants is that its rotation period is expected to be longer and its magnetic field is much weaker giving very weak Stokes $V$ signals. However the  ZDI method gave sound results in the case of EK\,Eri, which has $P_{\rm rot}$ = 309\,d, and in the case of \object{Vega,} which has a 1\,G magnetic field. Vega is even more difficult to detect since fewer spectral lines are available (Petit et al. 2010).

\subsection {Rotation period determination}
Using ZDI we inferred the rotation period $P_{\rm rot}$ and the magnetic topology of Pollux. We limited the number of spherical harmonics to $l < 10$ since increasing the limit did not significantly change the results; almost all the magnetic field is included in the spherical harmonics modes with $l < 3$.
 We followed the approach of Petit et al (2002) for determining the $P_{\rm rot}$ of Pollux, and Fig.\,3 shows the obtained periodogram. The deepest minimum at 660\,d is interpreted as corresponding to the $P_{\rm rot}$ of Pollux, with a statistical error of 15\,d. Our investigation does not detect either solar or antisolar differential rotation. One weak minimum at 365\,d is interpreted as an alias owing to the seasonal sampling. The 660\,d period is longer than the period of the RV variations (589.64\,d, Hatzes et al. 2006) and than the inferred $P_{\rm rot}=587.2$\,d provided in Auri\`ere et al. (2014) from ZDI. 
The likely erroneous $P_{\rm rot}$ presented in the preliminary work by Auri\`ere et al. (2014) is due to an initially insufficient number of digits in the coding of the rotational phase, in comparison to the very small changes of phase between consecutive spectra of a nightly series.

In addition to statistical errors, the 660\,d period found by ZDI can be affected by systematic errors. In the case of Pollux we might be concerned by changes in the magnetic configuration, since observations span over several years and more than two rotations. Differential rotation might also cause concern, although it is not detected by the ZDI process. Solar-like differential rotation might provide the longer period obtained from ZDI with Stokes $V$ data mainly sensitive to polar regions where the large-scale field is strongest as presented by our magnetic map (Sect.\,4.2 just below), whereas RV data are mostly sensitive to equatorial regions. The difference between 660\,d and 590\,d  would correspond to about 3\%  of the solar differential rotation or to a small change in latitude. A total error of a few tens of days on the $P_{\rm rot}$ would not change the conclusions of our magnetic study. On the other hand, the possibility of  equality between the rotation period and the RV period (590\,d) raises the hypothesis that the magnetic field could be the origin of RV variations. In this case, the magnetic activity  has to be stable during the 25-year span of the Hatzes et al. (2006) RV observations. During the two rotations, to check the stability
of our magnetic model presented in Sect.\,4.2, we divided the observations in two subsets that cover about one rotation each and we compared the two maps and models obtained with ZDI. The uncertainties on each parameter determination are increased and we do not find any significant change between the two models nor with the parameters obtained for the full data. In particular we get the $P_{\rm rot}$ values of 649\,d and 643\,d, lower but consistent with the 660\,d value and its statistical error of 15\,d. 

Finally, our ZDI work gives a value of about 660 d for the $P_{\rm rot}$ of Pollux, but because of a statistical error of 15\,d and possible systematic errors including differential rotation effects, we cannot exclude that it could be equal to the 590\,d period of the RV variations.

\subsection {Magnetic model}
With the new version of the ZDI code, $l < 10$ and $P_{\rm rot}$=660\,d, our preferred ZDI model is as follows. The magnetic field is found to be very axisymmetric (95\% of the total field).  The inclination angle $i$ optimizing the ZDI model is equal to 50$^\circ$. The poloidal and toroidal components contain 53\% and 47\%, respectively, of the reconstructed magnetic energy. About 95\% of the poloidal magnetic energy corresponds to the dipole component (2.3\% and 1.7\% corresponding to the quadrupole and octopole components, respectively).  Figure\,4 presents the ZDI map corresponding to our model.  Figure\,5 compares the observed Stokes $V$ and Stokes $I$ LSD profiles to the synthetic averaged profiles of our magnetic model.

This fraction of toroidal field energy may appear high for a weakly magnetic star (e.g., See et al. 2015 in the case of main-sequence stars). This toroidal component could be due to a redshift of about 570\,\ms\ between Stokes $V$ and Stokes $I$ profiles, which the ZDI algorithm interprets as a ring of azimuthal field. This redshift is illustrated in Fig.\,6, which compares the derivative of the averaged Stokes $I$ profile and the averaged Stokes $V$ profile. The resulting sub-gauss toroidal component would be negligible for more active stars, where surface large-scale magnetic fields can reach tens to hundreds of gauss. In the case of Pollux, however, the consequence is an apparent balance between the poloidal and toroidal field. This redshift of Stokes $V$ with respect to Stokes $I$ could suggest a link of magnetic features with convective downflows, as observed in \object{Betelgeuse} (Mathias et al. 2018). An alternate ZDI reconstruction, where the redshift of Stokes V has been corrected for, leads to a magnetic map shown in Fig.\,7 that is vastly dominated by the poloidal field component. The poloidal field contains 99.95 \% of the reconstructed magnetic energy  and is dipolar for about 97\%. The angle $\beta$ between the rotation axis and the magnetic dipole axis is 10.5$^\circ$. The mean surface magnetic field $B_{\rm mean}$ is 0.44\,G, the radial dipole strength is 0.8\,G and its longitude corresponds to phase 0.4 in Fig.\,7. The maximum of $|B_\ell|$ of the fitted sinusoid shown in Fig.\,2 occurs at rotational phase 0.65 in Fig.\,7.

 Taking into account the $P_{\rm rot}$ of 660\,d, $i$=50$^\circ$, a radius of 8.8 $R_{\odot}$ (Nordgren et al. 2001 and Hipparcos parallax) and within the solid rotation hypothesis, the $v\sin i $ of Pollux is 0.52\,\kms. This value for  $v\sin i $ is significantly smaller than those presented in the literature (e.g., Gray 2014, $v\sin i  = 1.70 \pm 0.2$\kms). This very small $v\sin i $ is difficult to measure, as shown in the investigation by Auri\`ere et al. (2009), and it would require very high spectral resolution observations to be detected. Using $v\sin i  = 1.7$\,\kms\ (Gray 2014) would lead to a $P_{\rm rot}$ of about 200\,d. As shown in Fig.\,3, this period would give a ZDI solution with a significantly worse reduced ${\chi}^2$.

\begin{figure}
\centering
\includegraphics[width=11 cm,angle=0] {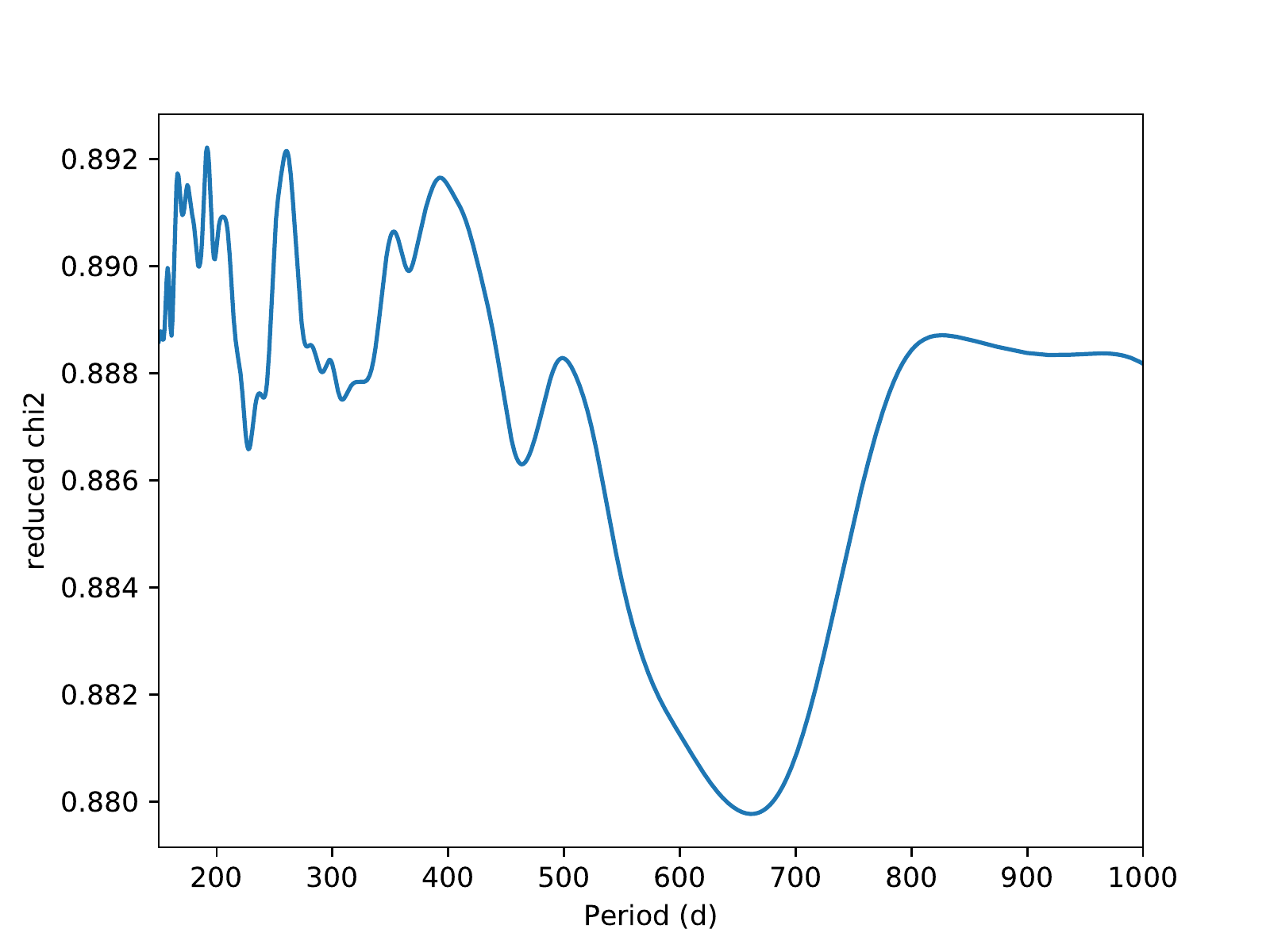} 

\caption{Periodogram of Pollux obtained using the Stokes $V$ spectra and the method of Petit et al. (2002) based on ZDI.}
\end{figure}

\begin{figure}
\centering
\includegraphics[width=9 cm,angle=0] {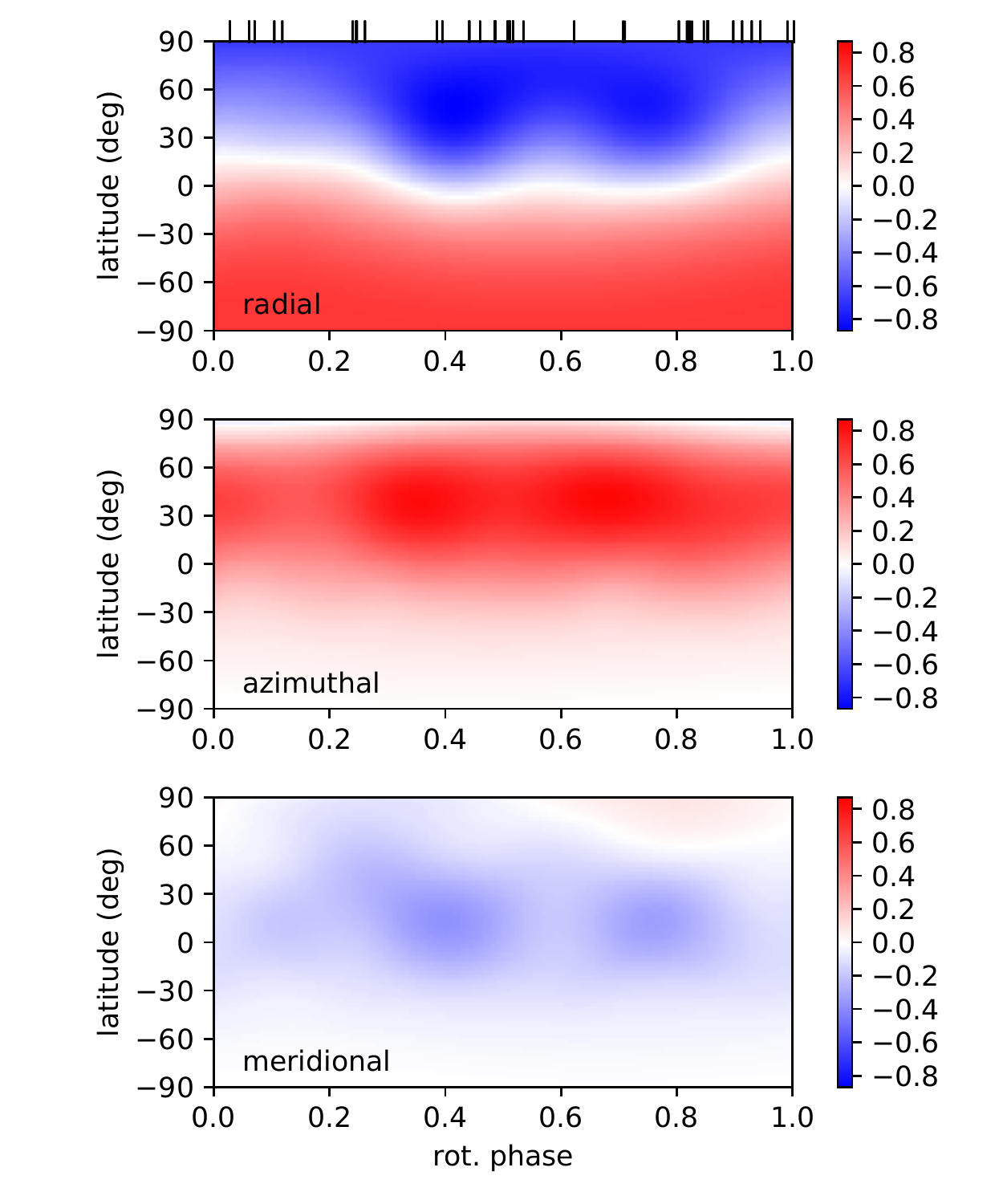} 

\caption{Zeeman Doppler image of Pollux. Each chart illustrates the field projection onto one axis of the spherical coordinate frame. The magnetic field strength is expressed in gauss. Ticks along the top chart indicate the observed rotational phases.}
\end{figure}  

\begin{figure}
\centering
\includegraphics[width=9 cm,angle=0]  {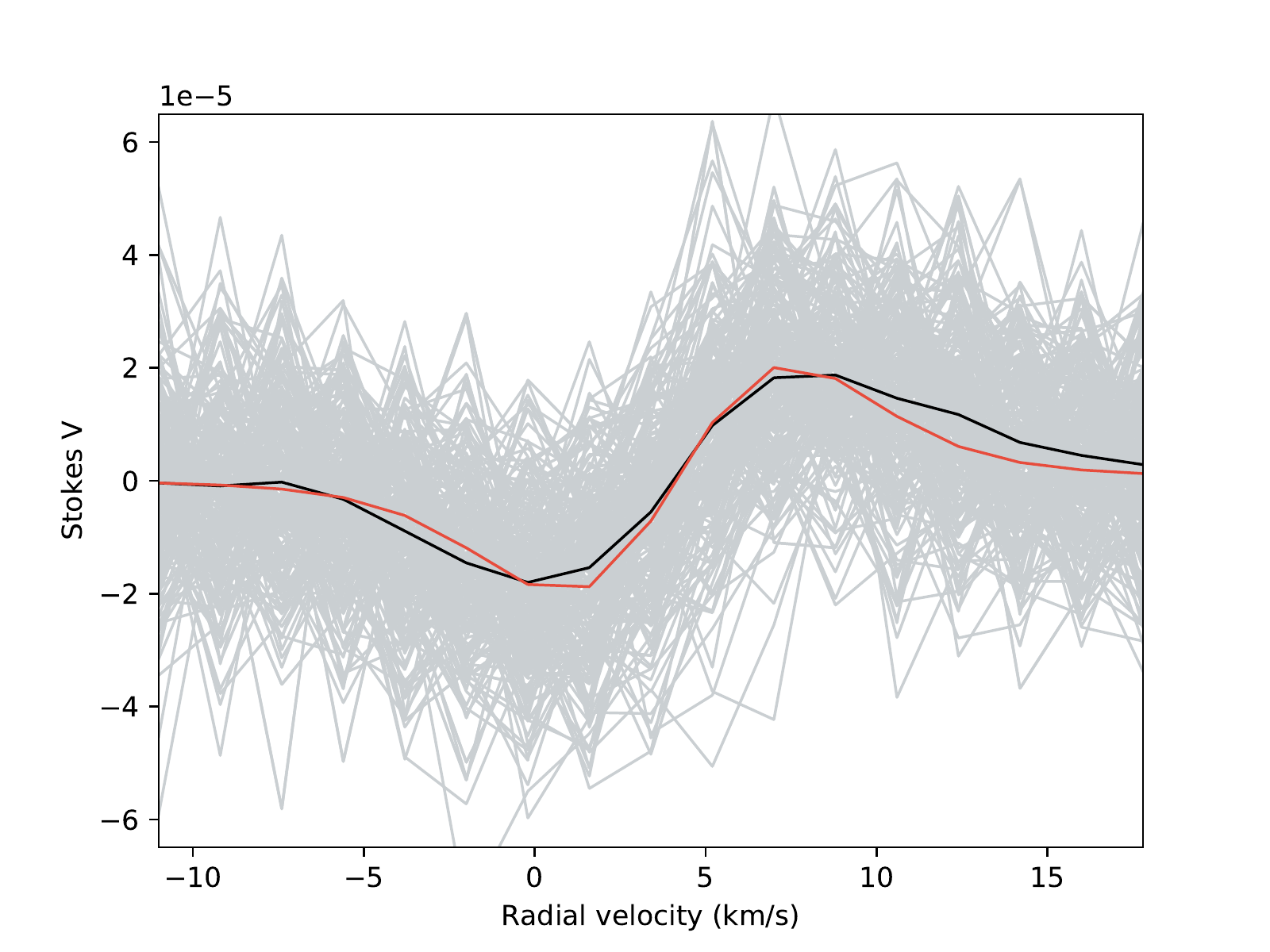}
\includegraphics[width=9 cm,angle=0]  {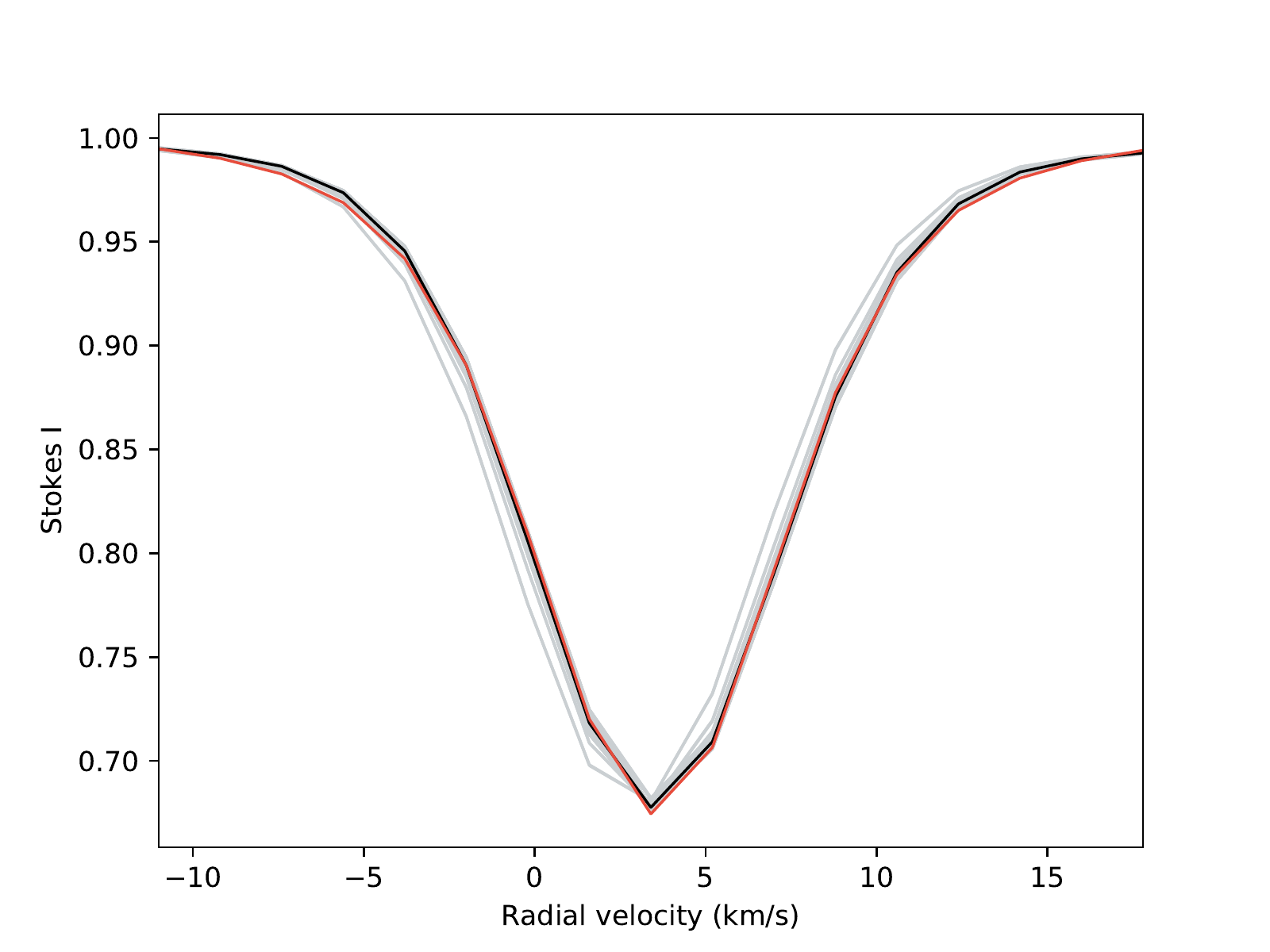}

\caption{Comparison between the observed  Stokes $V$ (upper plot) and Stokes I (lower plot) LSD profiles and the synthetic averaged profiles of our magnetic model. The X-axis is the RV and the ordinates are ($V/I_{\rm c}$) and ($I/I_{\rm c}$), respectively. The observations are illustrated in gray (individual profiles) and in black (averaged profiles), while the synthetic profiles are shown in red. }
\end{figure}  

\begin{figure}
\centering
\includegraphics[width=9 cm,angle=0] {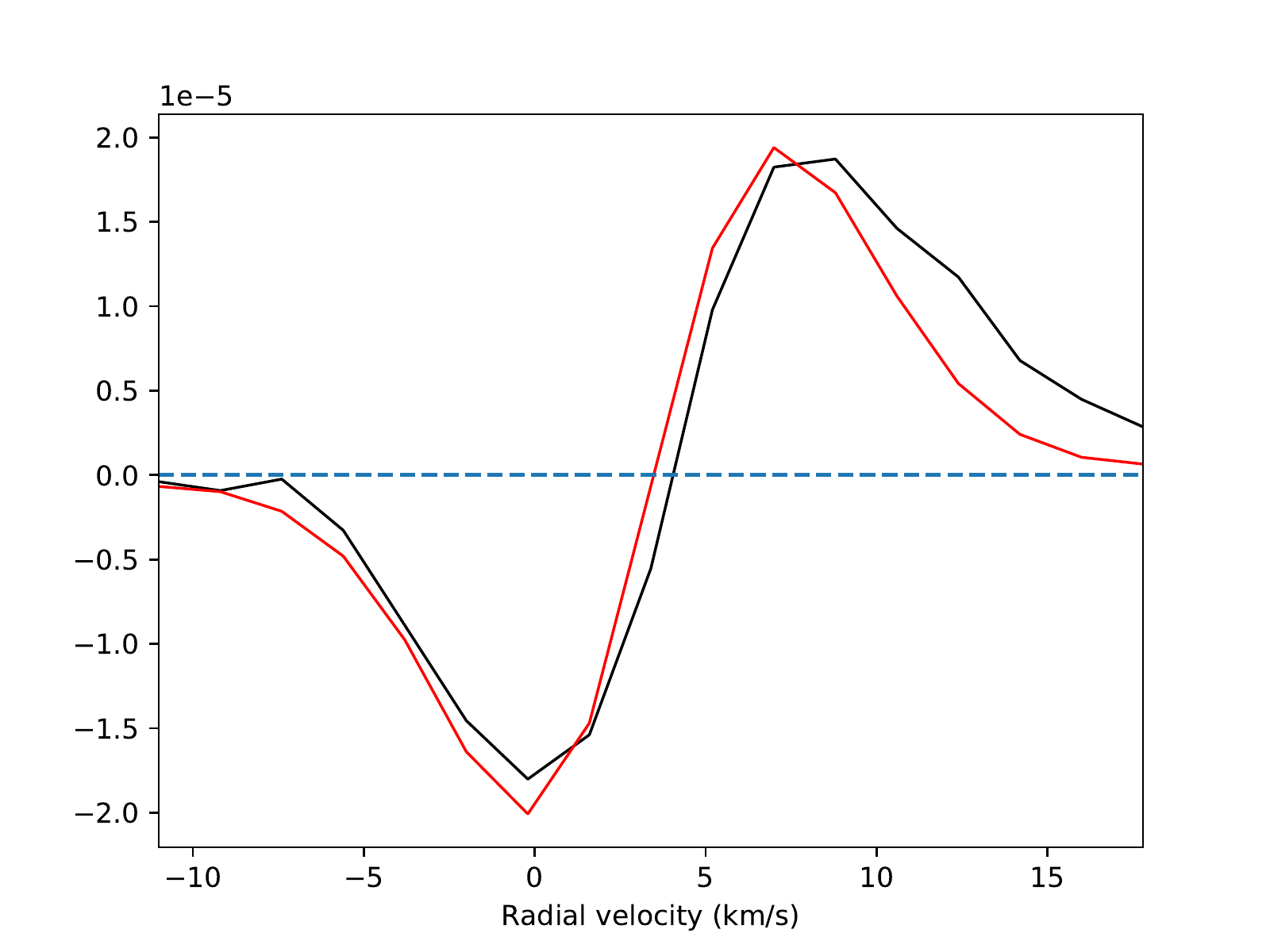} 
\caption{Derivative of the averaged Stokes $I$ profile (red) compared to the averaged Stokes $V$ profile (black). The Y-axis is scaled as $V/I_{\rm c}$.}
\end{figure}  

\begin{figure}
\centering
\includegraphics[width=9 cm,angle=0] {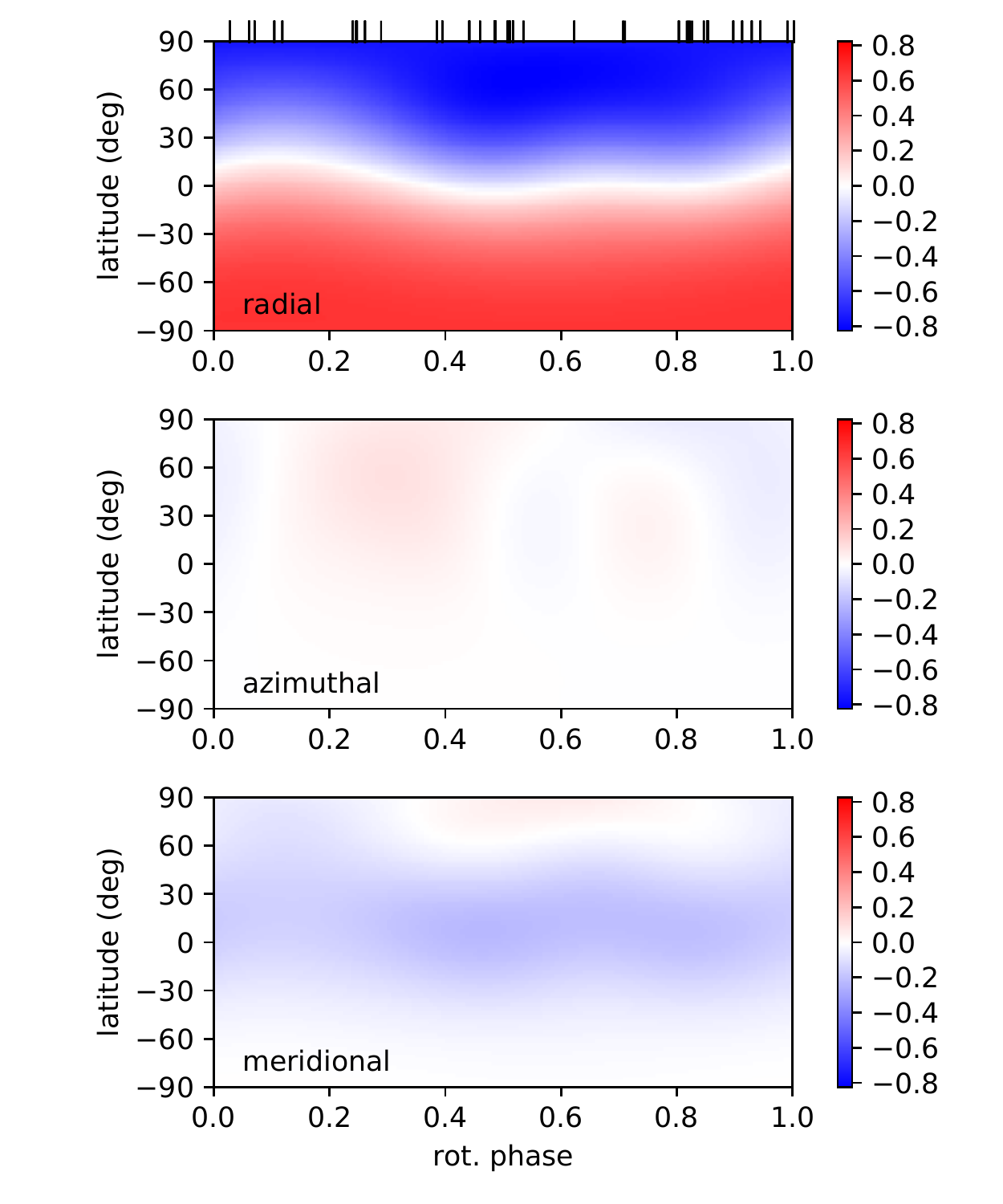} 
\caption{Zeeman Doppler image of Pollux with the Stokes $V$ profiles shifted by 570\,\ms\ as explained in Sect.\,4.2. Each chart illustrates the field projection onto one axis of the spherical coordinate frame. The magnetic field strength is expressed in gauss. Ticks along the top chart indicate the observed rotational phases.}
\end{figure}

\section {Origin of the magnetic field of Pollux}

 Pollux is the first red giant with a sub-gauss mean surface magnetic field for which ZDI is performed and $P_{\rm rot}$ is determined. This $P_{\rm rot}$  is also the longest determined up to now for a red giant (AKC).  The high Rossby number $\rm Ro$ of 2 inferred for Pollux using this $P_{\rm rot}$ and the convective turnover time $\tau_{\rm conv}$ = 330\,d derived using the evolutionary models with rotation of Lagarde et al. (2012) and Charbonnel et al. (2017) is the highest for the red giants with measured $P_{\rm rot}$ in the sample of AKC. The magnetic field origin of the other giants than Pollux  was suggested to be an $\alpha$ $\Omega$ dynamo or to descend from magnetic Ap stars (AKC).

\subsection {Ap star descendant hypothesis}

The high  $\rm Ro$, the dominant dipolar component of the reconstructed magnetic energy, and the stability of the magnetic field during two rotations would suggest that Pollux is an Ap star descendant. However using 1\,G for the dipole strength and taking into account the magnetic flux conservation hypothesis, we find that an Ap star progenitor of Pollux (taking a mass of 2.5\,$M_{\odot}$, AKC) would have a magnetic strength of about 20\,G, that is, far below the 100-300\,G lower limit found for an Ap star dipole strength by Auri\`ere  et al. (2007). Unlike the present Ap star descendant candidates, Pollux does not appear as a stronger magnetic star with respect to other red giants of its class; the Ap star descendant candidates
for which ZDI investigations have been done are listed Sect.\,4. 
 Therefore we consider that it is very unlikely that Pollux would be the descendant of a weak (and rare) magnetic Ap star.

\subsection {Dynamo operation hypothesis}

While it falls below the overall trends, Pollux is not a strong outlier in the plots presented by AKC showing the correlation of the magnetic field strength with $P_{\rm rot}$, $\rm Ro,$ and $S$-index for red giants harboring an $\alpha$ $\Omega$ dynamo-driven magnetic field.

The weak magnetic activity of Pollux compares well with that of the \object{Sun} observed as a star. The solar mean magnetic field (SMMF) representing the disk-averaged line-of-sight magnetic field of the Sun has a  strength varying between 0.15 and 1\,G with respect to the solar cycle (Garcia et al. 1999), when the $B_{\rm mean}$  of Pollux is about 0.4\,G (Sect.\,4) and  $|B_\ell|$ varies in the range 0.1-0.7\,G (Table\,1). The $S$-index of the global Sun varies during the solar cycle mainly in the range about 0.17-0.19 (Meunier 2018), that is, a little larger than that corresponding to the basal chromospheric flux. The  $S$-index of Pollux is stable at about 0.118 and very close to the value corresponding to the basal chromospheric flux of red giants (Schr\"oder et al. 2012). The average level X-ray luminosity of the Sun (Judge et al. 2003) is about that of Pollux (about $10^{27}$\,erg\,s$^{-1}$; Schr\"oder et al. 1998). This means that the magnetic activity of Pollux is similar to that of the Sun observed as a star, in which an $\alpha$ $\Omega$ dynamo is known to be involved. 

However the stellar parameters of Pollux are very different from those of the Sun; the $P_{\rm rot}$ is 20 times longer, the mass 2 times larger, and the radius is 9 times larger.
An extreme case of a dynamo when rotation is not involved is a local dynamo as inferred for the red supergiant Betelgeuse (Auri\`ere  et al. 2010). In the case of Betelgeuse, the magnetic field is observed to vary intrinsically on several timescales from weeks to months (Mathias et al. 2018) when we observe a 660\,d period, semi-sinusoidal variation for Pollux. Pollux is in mass and radius, intermediate between the Sun and Betelgeuse. Red giants and supergiants are expected to have giant convective cells with respect to the Sun (Schwarzschild 1975). The convective cells of Betelgeuse have been imaged  by spectropolarimetry and both  their upflow and downflow speeds are derived (Mathias et al. 2018, L\`opez Ariste et al. 2018). The upflows are traced by the bright center of convective cells and linear polarization, while the downflows of the same convective cells concentrate the observed magnetic field in circular polarization. Weak magnetic fields are detected, which are concentrated in the downflow lanes in between granules, similar to the quiet Sun magnetism. This suggests that this process, occurring in the quiet Sun and in Betelgeuse, is at work in Pollux in addition to a global component.

 Three-dimensional nonlinear magnetohydrodynamic (MHD) simulations were performed with the ASH code for a star representative of Pollux with $\Omega_{*} = \Omega_{\odot}/20$ (Palacios \& Brun 2014, Brun \& Palacios 2015). This study shows that this kind of slowly rotating red giant star is likely to possess global magnetic field maintained through contemporaneous dynamo action. The strength of the large-scale magnetic components is found to be on the order of a few gauss. These simulations are consistent with the results of this work  and suggest that Pollux is not a unique case of a weakly magnetic slowly rotating red giant.

A few other red giants have been detected with magnetic field strength at the sub-gauss or gauss level (AKC, Konstantinova-Antova et al. 2014). These stars are very similar in other respects to ordinary giants, with a small $S$-index, near the value corresponding to the  basal chromospheric flux.

Pollux is thus a representative of a class of slowly rotating and weakly magnetic red giants. Among these objects Pollux presents a  stability of its magnetic configuration during two rotations. As to the origin of this magnetic field, we favor the case of contemporaneous dynamo action in which rotation plays a role. Since Pollux falls below the overall trends in the plots presented in AKC for red giants harboring an $\alpha$ $\Omega$ dynamo-driven magnetic field the dynamo regime could be weaker; for example, turbulent dynamo as has been suggested to occur in asymptotic branch red giants (Soker \& Zoabi 2002), or  a universal turbulence-related dynamo mecanism explaining magnetic activity levels of both main sequence and evolved stars (Lehtinen et al. 2020). 

\section {Origin of the radial velocity variations of Pollux}

The variations of the RV of Pollux have been unanimously considered to be the result of a hosted planet (e.g., Hatzes et al. 2006, Reffert et al. 2006, Han et al. 2008). Table\,2 shows that the planet hosted by Pollux inferred by these authors  has a mass higher than about 2.5\,$M_{\rm Jup}$ (for a mass of Pollux of 1.7\,$M_{\odot}$ used in these three references) and an orbital semimajor axis of about $a = 1.64$\,AU.
We now have to consider how the weak magnetic field and the associated stellar activity could contribute to the RV variations of Pollux. 

\subsection {Surface magnetic field of Pollux: Incidence on RV measurements}

The effect of magnetic activity on RV measurements of stars has been investigated for a long time in the case of stellar spots or other  magnetic features (e.g., Hatzes 2002, H\'ebrard et al. 2014). However these studies involved magnetic fields that were generally stronger than a few tens or hundreds of G. In the context of the weak magnetic field of Pollux, since it is comparable to that of the Sun observed as a star (see Sect.\,5.2), it is relevant to refer to the solar studies made in the scope of disentangling weak activity effects and orbital effects on RV variations in the case of stars hosting planets (e.g., Meunier et al. 2010a,b, Haywood et al. 2016, Collier Cameron et al. 2019).  These  studies of the Sun observed as a star are based both on simulations and observations of the Sun from space or the ground. From these studies, several learnings can enlighten Pollux's case.

1) Space observations using SDO/HMI and SDO/AIA (e.g., Bose and Nagaraju 2018) or using MDI/SOHO (e.g., Meunier et al. 2010b) enabled us to disentangle contributions from different magnetic features to the global solar properties. Efficient  magnetic activity contributions to the RV variations observed on the global Sun are through attenuation of the convective blueshift (CB)  and by actual downflows (as linked with supergranules in the Sun or with the large convective cells in Betelgeuse). 

2) In the Sun as a star, variations of up to 10\,\ms\ are observed along the magnetic cycle (e.g., Meunier et al. 2010b). Variations of about 5\,\ms\ are linked to the transit of active  regions (Haywood et al. 2016, Collier Cameron et al. 2019). 
 
The solar example presented above shows that a main-sequence star with an activity level similar to that of Pollux can induce RV variations of a few \ms\ to 10\,\ms, when including both rotation and magnetic cycle effects.  The process at work in Pollux should not be due  to transit  of strongly magnetic structures such as solar spots or plages since no modulation near 600 days is observed on magnetic indicators apart from RV and Stokes $V$ data (Hatzes et al. 2006,  Gray 2014, this work). Pollux is more massive and more evolved than the Sun, but we noted in Sect.\,5.2 that some similar processes exist concerning weak magnetic fields between the Sun and the supergiant Betelgeuse, which could also occur in Pollux. This includes actual downflows, which could be linked to a redshift of Stokes $V$ profiles with respect to Stokes $I$, revealed by our ZDI investigation in Sect.\,4.2. 

Investigating the effect of the weak surface magnetic field of a red giant as Pollux on RV measurements is out of the scope of this work. The differences of atmospheric conditions at the surface of Pollux with respect to the Sun, for example lower gravity, are significant. The equipartition magnetic field is going to be smaller in Pollux and a 1\,G surface magnetic field is likely to have a much larger dynamical impact on the convection than in the Sun. We then  consider that the surface magnetic field of Pollux might contribute  significantly to the sinusoidal RV variations of 40\,\ms\ semi-amplitude.

\subsection {Scenarios explaining the radial velocity variations of Pollux}

\subsubsection {Only one significant period detected for the RV variations}
 A mixing of contributions from one hosted planet and from magnetic activity is expected to provide the periodic RV variations of Pollux. The observed period of 589.64\,d would be a combination of orbital and rotation periods. In this context we have to check if the two periods appear in the periodogram of the RV measurements. Hatzes et al. (2006) made a deep investigation in this direction using their long timescale sample of Pollux spectra. On their Lomb-Scargle periodogram of the residual RV measurements after subtracting the contribution of the orbital motion due to the companion the highest peak corresponds to a period of 121 days and is not considered as significant. However since it is near the 135\,d inferred for the photometric variations observed by Hipparcos, Hatzes et al. (2006) suggested that the rotation period  of Pollux would be about 130 days. From our knowledge of the magnetic red giants (AKC), this value is much too small for Pollux's $P_{\rm rot}$ since it would imply an active giant with a $|B_\ell|_{\max}$ that is five times larger than that observed and much larger $S$-index and $L_X$ activity indicators. Therefore we can conclude that only one period gives significant contribution to the very stable periodic
variations of RV, which is of about 590\,d. We then consider two scenarios. In each scenario, one effect dominates to produce the stable RV variations of Pollux.

\subsubsection {Planet hypothesis}
This corresponds to a general case in which the RV period is different from the $P_{\rm rot}$  and the magnetic activity of the star is weak. The planet is the main driver of the RV variations with a weak effect of the magnetic activity of Pollux, which cannot be disentangled from the present observations. 

\subsubsection {Magnetic activity hypothesis}
This is the peculiar case in which the observed period of the RV variations (about 590\,d) is the $P_{\rm rot}$ of Pollux, which cannot be discarded completely from our ZDI determination ($P_{\rm rot}$ about 660\,d, see Sect.\,4.1).  
Sect.\,6.1 shows that RV variations of about 10\,\ms\ are observed  in the solar case taken as an example of a star with a similar magnetic  activity level as that of Pollux. Then a planet could exist as well to fulfil the total
40\,\ms\ semi-amplitude RV variations. The revolution of the planet would be synchronized with the rotation of the star.  Because of the large semimajor axis and the long rotation period, the expected tidal effects appear very weak for  modifying the rotation period or the orbit (e.g., Pont 2009, Gallet et al. 2017), and the synchronous rotation would have to be explained. However the atmospheric conditions in  Pollux are  different from that in the Sun as argued in Sect.\,6.1, and we cannot discard the possibility that the activity of Pollux could explain the totality of the RV variations and that the planet hypothesis would appear unnecessary.

\section {Magnetic field and the RV variations in planet-hosting red giants other than Pollux}

About 112 credible substellar companion candidates orbiting 102 G and K giant stars have been found so far (Reichert et al. 2019; a link to a catalog is given in the introduction of the paper).
From this catalog, a very well-marked peak is seen at spectral type K0III, which is that of Pollux, and almost all stars are between K0 and K2. The study of these systems is of importance since it would enable us to explore the fate of planetary systems during stellar evolution, as well as to access to exoplanets around stars with higher mass than the Sun.

In addition to Pollux, three stars of the above list of credible candidates for hosting planets, of spectral types between G9III and K1III, have been observed with Narval or ESPaDOnS (AKC; Konstantinova-Antova et al. 2014) with a sub-gauss accuracy. Two of these stars, \object{$\iota$\,Dra} and \object{$\gamma$$_1$\,Leo} do not show any detectable Zeeman-signal and have small $S$-indices. This means that they do not host an observable magnetic field as Pollux does, and their magnetic  activity may not contribute significantly to the observed large RV variations (amplitude six and four times larger than in Pollux). On the other hand, the third, \object{$\epsilon$ Tau} is Zeeman-detected and its magnetic properties are found to be similar to those of Pollux; the data are still too scarce to enable any conclusion with respect to eventual correlations to RV (AKC).

 Recent studies show that for a number of  intermediate-mass red giants, the planet hypothesis inferred for explaining sinusoidal RV variations should be abandoned because the RV signal does not remain perfectly stable  when a long-term study is performed (Delgado Mena et al. 2018, for several open cluster stars; Hatzes et al. 2018, for \object{$\gamma$ Dra}; Reichert et al. 2019, for \object{Aldebaran}). The alternative  process (stellar oscillations?) is still unclear but it appears clear that we should be cautious in interpreting RV variations as due to companions in giant stars (Hatzes et al. 2015). 

 Aldebaran, as well as $\gamma$ Dra, are K5III and significantly cooler than Pollux. Sinusoidal variations of RV were discovered for Aldebaran by Hatzes et al. (1993) at the same time as for Pollux. Recently Hatzes et al. (2015) argued that these variations could be explained by a planetary companion of revolution period of 629\,d, with stellar activity in addition. From observations of activity indicators, these authors suggested that the rotation period of Aldebaran is 520\,d. Then,  Reichert et al. (2019) presented a new study  that does not support the planet hypothesis; oscillatory convective modes might be a plausible alternative explanation of the observed RV variations.

 The AKC work has Zeeman-detected Aldebaran. Aldebaran presents a variable weak magnetic field with a reversing $B_\ell$ and a variable $S$-index significantly larger than that of Pollux.  A correlation between RV variations and $B_\ell$ as well as activity indicators could not be concluded for Aldebaran from their sparse sample measurements. The results cited above suggest that RV variations coming from combined effects of a planet and magnetic activity could occur in the case of Aldebaran.

 A substantial fraction of the G-K red giants hosting planets are intermediate-mass stars and a large percentage of these stars could be weakly magnetic  (AKC; Konstantinova-Antova et al. 2014). Our preliminary investigation presented above concerns Pollux  and three other stars of about the same spectral type: Pollux and $\epsilon$ Tau present similar magnetic activity and $\iota$ Dra and $\gamma$$_1$ Leo did not present  magnetic activity during our observations. The magnetic activity would not significantly impact the planet-driven RV variations in the case of $\iota$ Dra and $\gamma$$_1$ Leo, and new observations are required to clarify the situation in the case of Pollux and $\epsilon$ Tau. In other cases, as possibly in Aldebaran, the observed RV variations could be a combination of orbital and rotational effects. The case of planet hosting red giant candidates with small amplitude RV variations particularly need to be checked for a magnetic activity contribution.

\section{Conclusions}

We have monitored the magnetic field of Pollux for a duration of 4.25 years, that is,  more than two times the period of the RV variations. The longitudinal magnetic field of Pollux is found to vary with a sinusoidal behavior and a period that is similar to that of the RV variations.  
Our ZDI investigation shows that  the rotation period of Pollux is equal to 660$\pm$15\,d, close to the RV variation period of 590\,d; the equality of the two periods is not excluded, particularly if we take into account possible systematic errors.  As to the magnetic topology, the poloidal component is dominant and almost purely dipolar, with an inclination of 10.5$^\circ$ of the dipole with respect to the rotation axis. 
The mean strength of the surface magnetic field is about 0.4\,G. We suggest that a weak-regime dynamo, as supported by 3-D nonlinear MHD simulations (Brun \& Palacios 2015), is the origin of the magnetic field of Pollux. In this way Pollux appears as the representative of a detected class of slowly rotating and  weakly magnetic G-K giants (AKC; Konstantinova-Antova et al. 2014). 
  
Pollux is about as active as the Sun observed as a star and this activity induces RV variations of a few \ms\ to about 10\,\ms\ for the global Sun when both rotation and magnetic cycles are taken into account. We suggest in Sect.\,6. that  the magnetic activity of the more massive and more evolved star Pollux could induce significant RV variations as well, which would be modulated by the rotation period. Since from the long-term study of Hatzes et al. (2006), only one period emerges to produce the stable sine shape of RV observations of Pollux, we propose two scenarios to explain these RV observations. If the RV period is different from the rotation period of Pollux, the observed periodic RV variations are due to the hosted planet and the contribution of Pollux magnetic activity is not significantly detected. In the peculiar case when the two periods are equal, the weak magnetic field would contribute to the RV variations and we cannot discard the possibility that the activity of Pollux could explain the total RV variations and that the planet hypothesis would appear unnecessary.

In any case,  magnetic activity could contribute significantly to RV variations in some intermediate-mass G-K red giants hosting planets, particularly those with small amplitude RV variations. Therefore new Zeeman observations of a number of red giant candidates for hosting planets (including Pollux) would be desirable for disentangling the contributions of orbital and rotation periods.

\begin{acknowledgements}
       We thank  the TBL and CFHT teams for providing service observing. We acknowledge financial support from 'Programme National de Physique Stellaire' (PNPS) of CNRS/INSU, France. JFD acknowledges funding from the European Research Council under the H2020 research \& innovation programme (grants $\#$740651 NewWorlds). GAW acknowledges support from the Natural Science and Engineering Research Council of Canada (NSERC).
\end{acknowledgements}

\end{document}